\documentclass[pdflatex,iicol,sn-aps]{sn-jnl}

\usepackage{graphicx}%
\usepackage{physics}
\usepackage{multirow}%
\usepackage{amsmath,amssymb,amsfonts}%
\usepackage{amsthm}%
\usepackage{mathrsfs}%
\usepackage[title]{appendix}%
\usepackage{xcolor}%
\usepackage{textcomp}%
\usepackage{manyfoot}%
\usepackage{booktabs}%
\usepackage{algorithm}%
\usepackage{algorithmicx}%
\usepackage{algpseudocode}%
\usepackage{listings}%
\usepackage{titlesec}

\theoremstyle{thmstyleone}%
\theoremstyle{thmstyletwo}%

\theoremstyle{thmstylethree}%

\begin{document}

\title[Article Title]{Transport in Single Quantum Dots: A Review from Linear Response to Nonlinear Regimes}


\author[1]{\fnm{Gustavo} \sur{Diniz}}

\author[2]{\fnm{Silvio} \sur{Quintino}}

\author[2]{\fnm{Vivian} \sur{V. França}}

\affil[1]{\small{São Paulo University (USP), São Carlos Institute of Physics, São Carlos}}

\affil[2]{\small{São Paulo State University (UNESP), Institute of Chemistry, Araraquara}}


\abstract{
Quantum dots are versatile systems for exploring quantum transport, electron correlations, and many-body phenomena such as the Kondo effect. While equilibrium properties are well understood through methods like the numerical renormalization group and density matrix renormalization group, nonequilibrium transport remains a major theoretical challenge. From the experimental point of view, recent advances in nanofabrication and measurement techniques have enabled the investigation of far-from-equilibrium regimes. These conditions give rise to new transport phenomena, where strong correlations and nonequilibrium dynamics interplay in complex ways --- beyond the reach of conventional linear response theory. To meet these challenges, new approaches such as nonequilibrium Green’s functions, real-time NRG, and time-dependent DMRG have emerged. This work reviews the established results for quantum dot transport in and beyond the linear regime, highlights recent theoretical and experimental advances, and discusses open problems and future prospects.}


\maketitle

\titleformat*{\section}{\large\bfseries}
\titleformat*{\subsection}{\normalsize\bfseries}

\section{Introduction}\label{sec1}

Understanding the transport properties of correlated electronic systems has been crucial for advancing new, more efficient electronic devices and, by extension, modern technologies \cite{RevModPhys.79.1217,PhysRevLett.61.2472,RevModPhys.64.849,Newns1998,RevModPhys.73.357}. Within this context, quantum dots have established themselves as versatile platforms for investigating fundamental aspects of quantum transport, electron-electron interactions, and many-body phenomena such as Kondo physics \cite{Pustilnik2004,PhysRevLett.90.076804,PhysRevLett.81.5225,PhysRevB.101.125115,PhysRevB.80.235317}.

Over the past three decades, much of the research on quantum dots has focused on the linear response regime (LRR) \cite{PhysRevB.80.235318,PhysRevB.80.235317,PhysRevB.106.075129,Pinto_2014}, where applied biases and temperature differences remain small, allowing transport properties to be described by equilibrium correlation functions via linear response theory (LRT) \cite{Kubo1957,Mahan2010-xj,PhysRevLett.68.2512}, since the system response is proportional to the applied external perturbation. LRT has proven extremely useful for studying conductance in these devices, as it offers a simple and inexpensive framework for computing their transport properties \cite{PhysRevB.80.235318,PhysRevB.80.235317,PhysRevB.106.075129,Pinto_2014}. Notably, predictions based on LRT have been experimentally verified across a wide range of geometries and parameters \cite{Konig1996,Xu2021,See2010,PhysRevLett.86.878,PhysRevB.83.115323,tenKate2022}.

However, recent technological advances in nanofabrication and measurement techniques have made it possible to explore transport regimes far beyond linear response, where large bias voltages, temperature gradients, and time-dependent perturbations drive the system out of equilibrium, manifesting entirely new transport dynamics that transcend established paradigms \cite{4408808,960386,PhysRevB.79.235336,PhysRevB.79.165413,PhysRevB.84.195116,Roch2008,Paaske2006,PhysRevLett.131.206303}. In this regime, strongly correlated effects, such as nonequilibrium Kondo physics \cite{PhysRevB.84.195116,Roch2008,Paaske2006}, and Coulomb blockade \cite{PhysRevLett.131.206303,PhysRevB.83.115323}, manifest in nontrivial ways, and conventional LRT fails to capture the complex interplay between interactions and nonequilibrium dynamics.

From a theoretical standpoint, techniques such as the numerical renormalization group (NRG) \cite{RevModPhys.47.773,RevModPhys.80.395} and the density matrix renormalization group (DMRG) \cite{PhysRevLett.69.2863,PhysRevLett.93.076401} have been instrumental in studying thermal equilibrium properties, as well as transport and excitation behavior with LRT. However, exploring out-of-equilibrium properties presents even greater challenges, as not only the ground state and low-energy excitations but also high-energy excitations can play a significant role. As a result, the nonequilibrium properties of these systems, in contrast to their equilibrium behavior, are still not fully understood \cite{PhysRevB.74.085324,PhysRevLett.101.166401,RevModPhys.83.863,PhysRevLett.130.106902}.

The experimental advances have further motivated theoretical efforts to improve nonequilibrium methods. New approaches such as nonequilibrium Green’s functions \cite{Do_2014,PhysRevB.74.085324}, real-time NRG \cite{PhysRevLett.95.196801,PhysRevB.98.155107,PhysRevB.89.075118,PhysRevB.74.245113,Picoli}, and time-dependent DMRG  \cite{PhysRevLett.93.076401,PhysRevB.79.235336} simulations, have significantly advanced our understanding of transport properties in quantum dots under finite bias, magnetic fields, and temperature differences.

Although intensively studied for decades, transport properties in quantum dots still uncover rich and not yet fully understood physical behavior. In particular, the investigation of transport phenomena beyond the LRR has uncovered a frontier of new effects. These studies are promising not only for advancing the fundamental understanding of nonequilibrium many-body systems \cite{PhysRevB.111.035445}, but also for enabling novel applications in electronic devices \cite{Liu2021} and quantum dot-based quantum computing \cite{PhysRevA.57.120}.

Here, we aim to provide an overview of the established properties of quantum dots and their transport behavior in the linear regime, summarize recent theoretical and experimental advances both within and beyond this regime, and discuss emerging perspectives while highlighting key open problems in the field.



\section{Quantum dot properties}

\begin{figure}[htb!]
	\centering
        \includegraphics[scale=0.33]{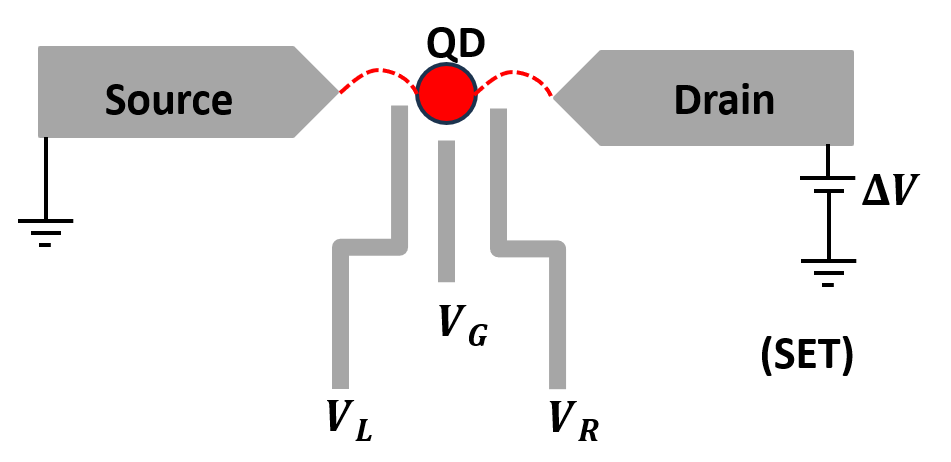}
        \includegraphics[scale=0.33]{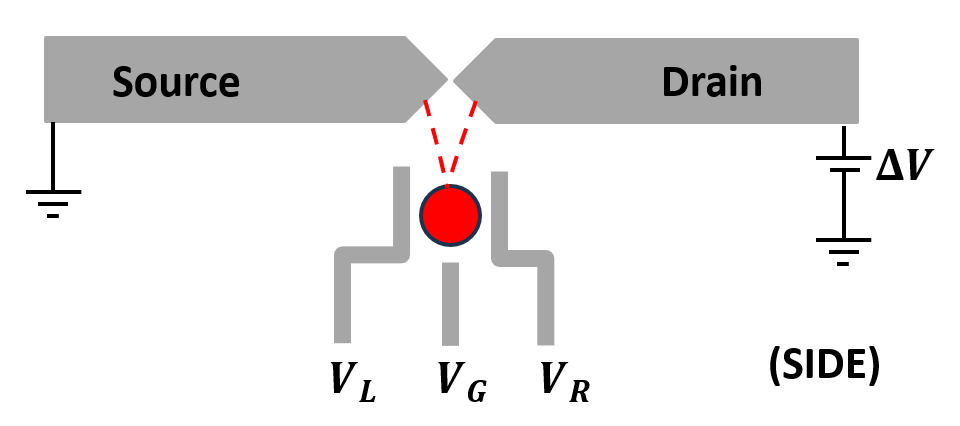}
	\caption{Schematic of QDs for SET (top) and SIDE (bottom) geometries. While in the SET geometry, electrons necessarily cross the QD to move from the source to the drain, in the SIDE, electrons can bypass the QD and cross directly. The electrodes $V_G$, $V_L$, and $V_R$, control the energy levels within the QD, the Coulomb repulsion, and its coupling to the metallic leads. Applying a small potential difference $\Delta V$ between the two leads enables transport through the devices.}
	\label{QDD}
\end{figure}

A quantum dot (QD) is a nanoscale structure in which electrons are confined in all spatial dimensions \cite{RevModPhys.79.1217}, leading to a discrete energy spectrum resembling that of an artificial atom. Typically, a QD is coupled to two leads: a source and a drain. Among the various QD configurations, two geometries are particularly noteworthy, as illustrated in Fig. \ref{QDD}: the single-electron transistor (SET) geometry \cite{PhysRevB.80.235317} and the side-coupled (SIDE) geometry \cite{PhysRevB.80.235318}. Thanks to their high tunability, QDs offer precise control over their electronic properties, making them ideal platforms for exploring transport phenomena in correlated many-body systems \cite{PhysRevLett.90.076804,PhysRevLett.81.5225,PhysRevB.101.125115,PhysRevB.80.235317}.

To understand the behavior of a QD, we begin with the single impurity Anderson model \cite{PhysRev.124.41}, described by the Hamiltonian
\begin{eqnarray}\label{AM}
 H_{0}=\varepsilon_d \left(n_{d\uparrow} + n_{d\downarrow} \right) + U n_{d\uparrow}n_{d\downarrow} 
+ H_B + H_{\mathrm{h}},
\end{eqnarray}
where $n_{d\sigma} = d_\sigma^\dagger d_\sigma$ is the electronic density, and $d_{\sigma}^\dagger$ ($d_{\sigma}$) creates (annihilates) one electron on the QD with spin $\sigma = \{ \uparrow, \downarrow \}$ and energy $\varepsilon_d$. The Coulomb repulsion term $U$ penalizes double occupancy. $H_B$ represents the leads, and $H_{\mathrm{h}}$ allows electron transfer between the dot and leads.

The conductor leads can be simply modeled by a one-dimensional tight-binding model as:
\begin{equation}\label{TBM}
\begin{aligned}
{H}_{B} = -\tau \sum_r \sum_{j=1, \sigma}^{N} \left(a_{r,j,\sigma}^{\dagger}a_{r,j+1,\sigma}+\mathrm{h.c.}\right).
\end{aligned}
\end{equation}
Here, $\tau$ is the hopping, $r = \{L,R\}$ and indicates the source (L) or the drain (R), the operator $a_{r,j}^{\dagger}$ creates one electron in the site $j$ of the lead $r$, and $N$ is the number of sites\footnote{In the case of the side-coupled geometry, the term $-\tau \sum_\sigma\left( a_{L,1,\sigma}^{\dagger}a_{R,1,\sigma} +\mathrm{h.c.}\right)$ has to be add in Eq. \eqref{TBM}, enabling electrons to transfer between the source and drain.}.

The hybridization term  can be written as:
\begin{eqnarray}\label{Hyb.RS}
 H_{\mathrm{h}} = \sqrt\frac{\Gamma}{\pi \rho_0} \sum_r \sum_\sigma \left( a_{r,1,\sigma}^\dagger d_\sigma + d^\dagger_\sigma a_{r,1,\sigma} \right),
\end{eqnarray}
where $\rho_0 = 1/2\pi\tau$ is the leads density of state at Fermi level, and $\Gamma$ is the hybridization strength.

The transport in the chain is initiated by introducing a small potential difference $ \Delta V $ between the source and the drain, which alters their respective chemical potentials, driving the QD out of equilibrium. This can be modeled by the term:
\begin{eqnarray}\label{Hddp}
\Delta H =\frac{e\,\Delta V}{2} \sum_{j \sigma} \left( a_{L,j\sigma}^\dagger a_{L,j\sigma} - a_{R,j\sigma}^\dagger a_{R,j\sigma} \right),
\end{eqnarray}
with $e$ being the positron charge.

\subsection{Equilibrium properties}

Among the various intriguing phenomena observed in quantum dots, Kondo physics stands out as a particularly rich one \cite{PhysRevB.80.235318,PhysRevB.106.075129,PhysRevB.101.125115,PhysRevB.80.235317}, emerging at low temperatures. Therefore, to extract the relevant physics, it requires an accurate description of the low-energy states. Since the Hamiltonian described by \eqref{AM} is not quadratic if $U \ne 0$, a closed exact analytical expression for this problem is not possible by conventional methods \cite{RevModPhys.47.773,RevModPhys.80.395}. To make the situation even worse, this problem is well known to be non-perturbative under the usual perturbative expansions \cite{RevModPhys.47.773,RevModPhys.80.395}, thus requiring an \emph{impurity solver} \cite{Pavarini2021} to accurately diagonalize the model and compute its properties.

Despite the challenges involved in diagonalizing the Hamiltonian \eqref{AM}, its equilibrium properties are well established and can be qualitatively described in terms of its \emph{fixed points} \cite{RevModPhys.47.773,RevModPhys.80.395,Nozieres1974_Kondo,Nozieres1978}. At high temperatures, $k_B T \gg \{ |\varepsilon_d|, |2\varepsilon_d + U| \}$, where $k_B$ is the Boltzmann constant, the system is in the \emph{free orbital} fixed point, such that the QD is effectively decoupled from the band. As the temperature decreases, for $\varepsilon_d < 0$, the QD begins to interact with the band, favoring the single occupancy in the dot. The QD then acquires an effective magnetic moment, a \emph{local moment} fixed point, and the QD starts to interact antiferromagnetically with the conduction electrons.

\begin{figure}[ht!]
    \centering
    \includegraphics[scale=0.43]{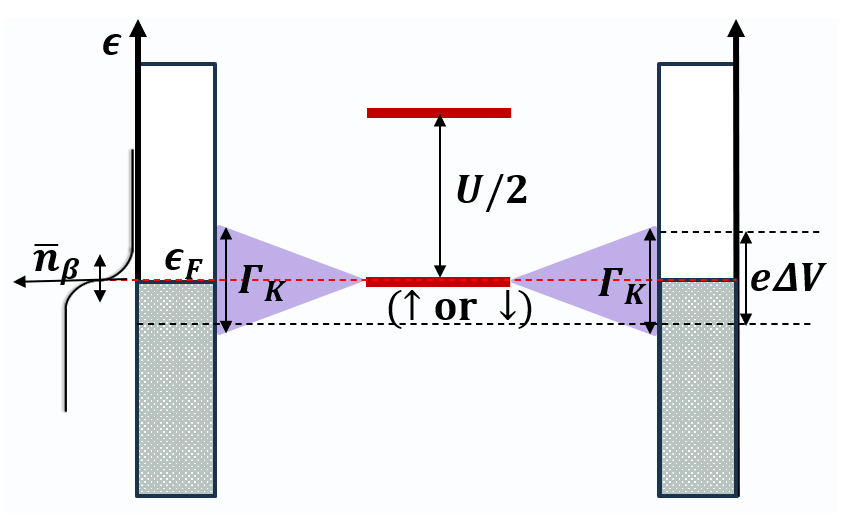}
    \caption{Schematics of the the Kondo hybridization ($\Gamma_K \sim k_B T_K$) between the QD and the low-energy states of the band to form the spin singlet. Once in the ground state, the QD is occupied by a single electron; to access double occupancy and conduct one electron, the Coulomb blockade $\sim U/2$ must be overcome. The high density of states around the Fermi energy $\epsilon_F$ enables conductance between the leads when $\Delta V \ne 0$. Here, $\bar n_\beta$ is the Fermi function.}
    \label{Kondo1}
\end{figure}

As the temperature decreases further, the antiferromagnetic coupling effectively increases. At a critical temperature, known as the \emph{Kondo temperature} $T_K$, the QD becomes strongly coupled to the band. This temperature characterizes the \emph{strong-coupling} fixed point and can be expressed by Haldane's formula \cite{PhysRevLett.40.416},
\begin{align}
    k_B T_K \approx 2\tau\sqrt \frac{2\Gamma}{\pi U} \exp( -\frac{\pi |\varepsilon_d||\varepsilon_d + U|}{4\Gamma U} ).
\end{align}

Below $T_K$, the dot and the neighboring sites, bound by an antiferromagnetic interaction \cite{RevModPhys.47.773}, form a spin-singlet many-body state, and the dot’s magnetic moment is fully screened by the conduction electrons near the Fermi level. This phenomenon gives rise to the well-known \emph{Kondo cloud} \cite{RevModPhys.47.773,Borzenets2020,PhysRevLett.110.246603,Prueser2011},  which spreads spatially within the leads and can, in principle, reach macroscopic sizes.

This singlet state arises from the QD hybridization with the low-energy levels of the band, with a strength given by $\Gamma_K \sim k_B T_K$ \cite{PhysRevB.80.235317}, as shown in the bottom panel of Fig. \ref{Kondo_2}. This interaction creates a high local density of states concentrated at the Fermi level, known as the \emph{Kondo resonance} \cite{PhysRevB.80.235317} --- a sharp peak at zero energy, shown in the top panel of Fig. \ref{Kondo_2}. The Kondo resonance can overcome the Coulomb blockade, enabling ballistic conduction through the SET QD (see bottom panel of Fig. \ref{Kondo_2}), inducing the phenomenon known as the \emph{zero-bias conductance anomaly} \cite{PhysRevB.80.235317}.

\subsection{Transport properties via LRT}

When the external perturbation is sufficiently small, one can rely on standard LRT \cite{Mahan2010-xj}, thereby avoiding the heavy computational cost of calculating nonequilibrium properties. In this formalism, as shown in Refs. \cite{PhysRevB.80.235317,Pinto_2014}, the zero-bias conductance for the SET geometry can be obtained via the Meir-Wingreen formula \cite{PhysRevLett.68.2512},
\begin{align}\label{GSET}
    G_{SET}(T)= G_0 \int d\epsilon \left( - \frac{\partial\bar n_\beta(\epsilon)}{\partial \epsilon} \right) \rho_d(\epsilon,T),
\end{align}
where $\bar n_\beta(\epsilon) = (1+\exp(\beta \epsilon) )^{-1}$ is the Fermi function, $\beta = 1/(k_B T)$, $G_0 = 2e^2/h$ is the conductance quanta and $h$ is the Planck constant. Here, $\rho_d(\epsilon,\beta)$ is the QD local density of states (LDOS) at finite temperature, which can be expressed as
\begin{align}\label{rhod}
    \rho_d(\epsilon,T) = \frac{\pi\Gamma}{\mathcal Z_\beta} &\sum_{n,m,\sigma}  \left( e^{-\beta E_n} + e^{-\beta E_m}   \right )\times \nonumber \\ &~|\bra {n} d^\dagger_\sigma\ket {m} |^2 \delta(\epsilon - E_n + E_m ),
\end{align}
where $\mathcal{Z_\beta} = \mathrm{Tr}\left[ \exp(-\beta H_0) \right] $, and $\ket{n}, \ket{m}$ represent many-body eigenstates of $H_0$, with energies $E_n$ and $E_m$ respectively.

As shown in Ref. \cite{Pinto_2014}, the SIDE and SET conductances are related by
\begin{align}\label{GSIDE}
G_{\mathrm{SIDE}}(T) = G_0 - G_{\mathrm{SET}}(T),
\end{align}
which means no additional computation is required: one can directly obtain $G_{SIDE}(T)$ from the SET conductance. Similarly, the thermal conductivity can be computed using the Wiedemann–Franz law \cite{Craven2020}.

\begin{figure}[ht!]
	\begin{center}
    \includegraphics[scale=0.41]{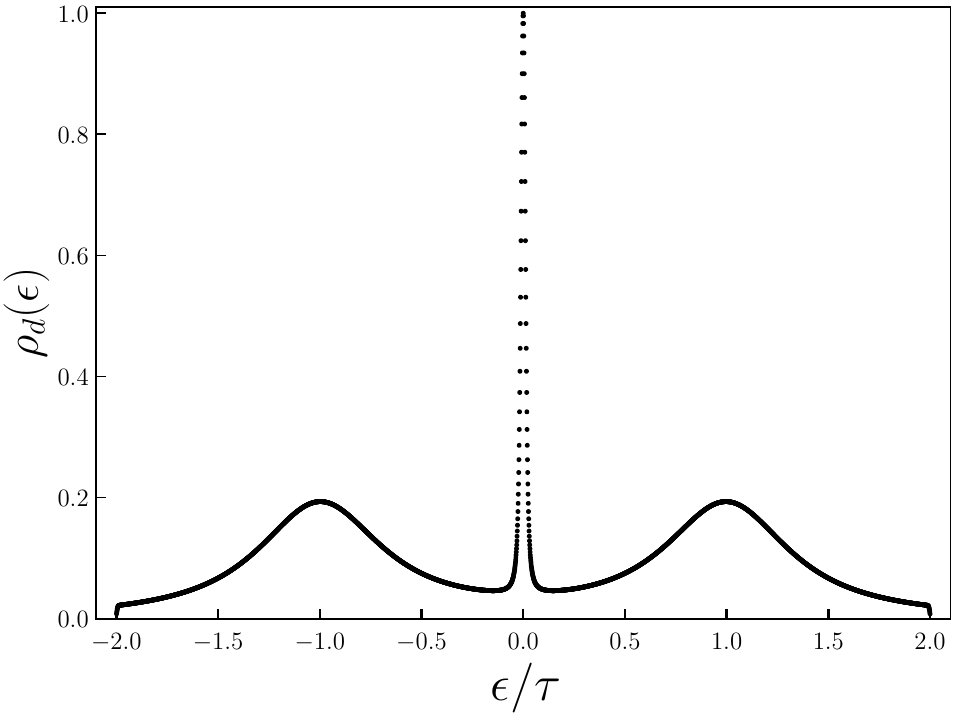}
     \includegraphics[scale=0.425]{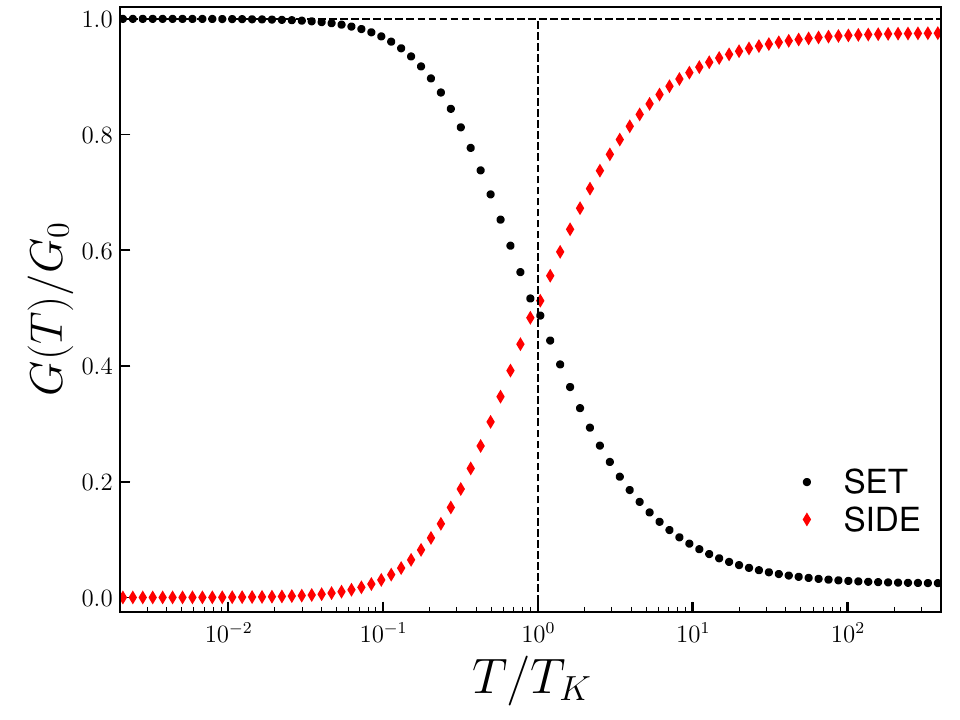}
    \end{center}
    \caption{(Top) LDOS as a function of energy $\epsilon$ at a fixed $k_B T = 10^{-3} \tau$ for the particle-hole symmetric case $\varepsilon_d = -U/2 = -\tau$, with hybridization $\Gamma = 0.20 \tau$. From the numerical data is extracted $k_B T_K \approx 0.04 \tau$ for this set of parameters. (Bottom) Zero-bias conductance as a function of temperature $T$, computed using Eq. \eqref{GSET} (SET) and Eq. \eqref{GSIDE} (SIDE). These quantities were calculated using the NRG method.}\label{Kondo_2}
\end{figure}


Figure \ref{Kondo_2} (bottom panel) shows the zero-bias conductances for the SET (via Eq. \eqref{GSET}) and the side-coupled (Eq. \eqref{GSIDE}) geometries. Both conductances were computed using the LDOS shown in the top panel. For the SET device, at low temperatures $T < T_K$, the QD is strongly coupled to the leads, the Kondo resonance appears in the LDOS, and these low-energy states provide a channel for electrons to transfer between the leads. For $T \gg T_K$, however, the QD-band coupling weakens, the QD can fluctuate into triplet states, the resonance peak broadens, the low-energy density at the QD decreases, and so does the conductance. The same Kondo physics occurs in the SIDE device; however, the direct conduction channel between the leads guarantees conductance when the QD-band channel is closed $T \gg T_K$. At low temperatures $T < T_K$ , the bands are strongly coupled to the QD, trapping the conduction electrons into the Kondo singlet, and the conductance drops to zero.

\begin{figure}[ht!]
	\begin{center}
    \includegraphics[scale=0.41]{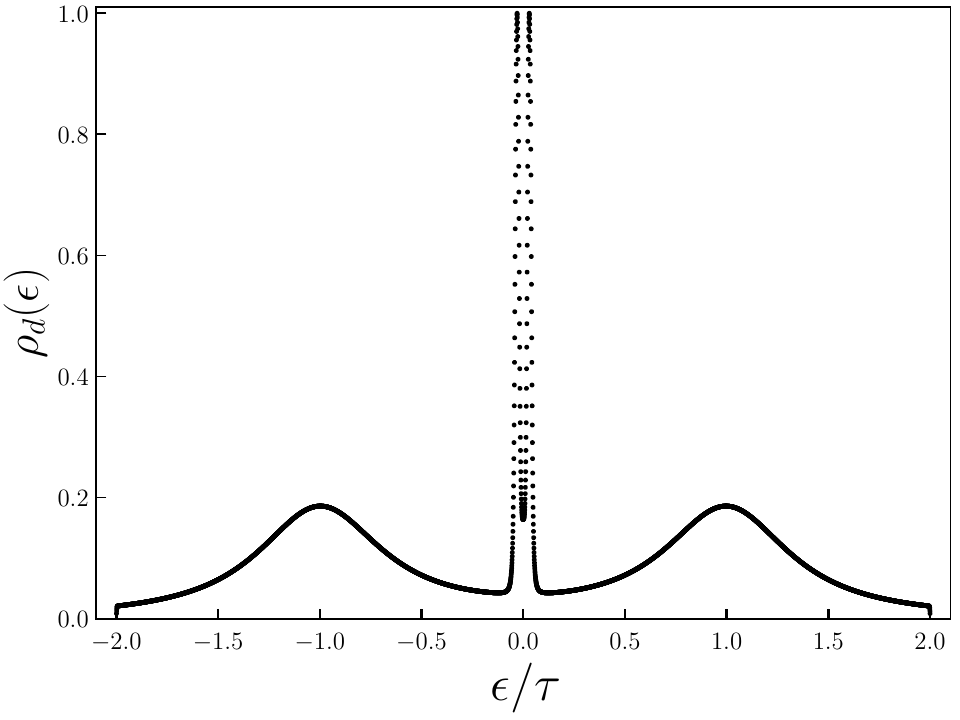}
     \includegraphics[scale=0.43]{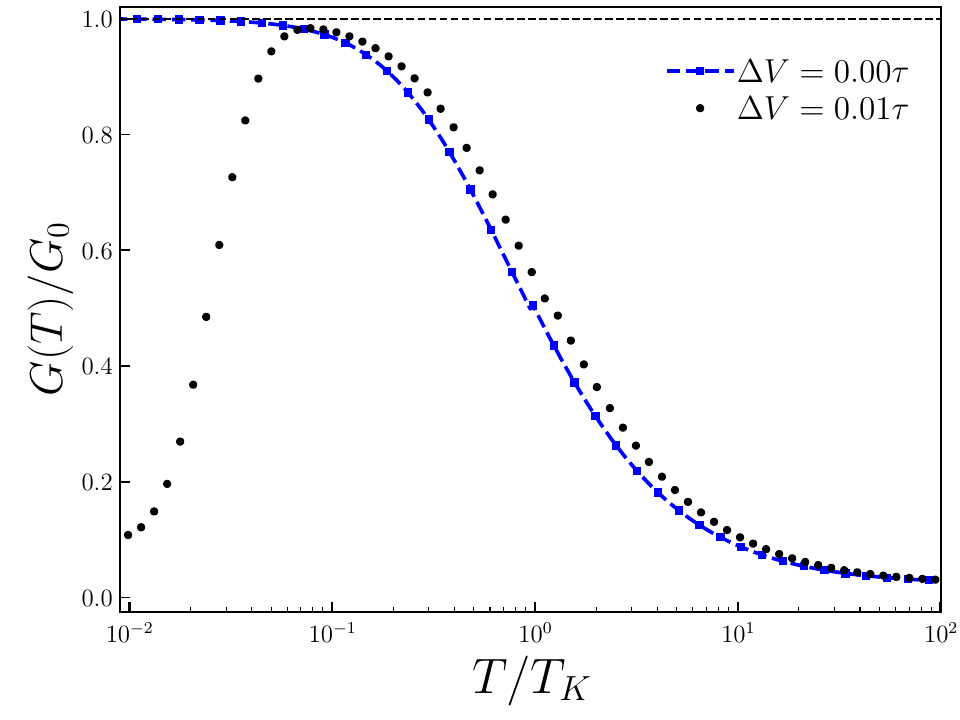}
    \end{center}
    \caption{(Top) LDOS as a function of energy $\epsilon$ at a fixed $k_B T = 10^{-3} \tau$ and a bias voltage $\Delta V = 0.01 \tau$, which causes a splitting of the Kondo resonance. The other parameters used here are the same as in Fig. \ref{Kondo_2}. (Bottom) Conductance for the SET geometry as a function of temperature, shown for finite (black dots) and zero bias (blue line). These quantities were calculated using the NRG method.}
	\label{Kondo_3}
\end{figure}

One notable effect is the splitting of the LDOS Kondo peak in the presence of a small non-zero voltage bias or a weak magnetic field \cite{Niu2015,PhysRevB.84.195116,Trocha2024,PhysRevB.106.125413}. Figure \ref{Kondo_3} shows this phenomenon (top panel) and its effect on the conductance (bottom panel): $G(T)$ deviates from the zero-bias behavior for $T<<T_K$, where it decreases due to bias-induced changes in the chemical potential \cite{PhysRevB.64.153305}. In this regime, the lowest energy states are more localized in one lead, and transferring an electron to the other requires overcoming the energy splitting \cite{PhysRevB.64.153305}. At higher temperatures, the behavior remains similar to the zero-bias case.


LRT has been extremely useful for studying the conductance in QDs, as it offers a simple and inexpensive platform to compute their properties and shows good agreement with experiments within the LRR \cite{Konig1996,Xu2021,See2010,PhysRevLett.86.878,PhysRevB.83.115323,tenKate2022}. However, there are situations where traditional LRT cannot be applied.

\begin{figure*}[ht!]
	\begin{center}
        \begin{tabular}{cc}
        \includegraphics[scale=0.418]{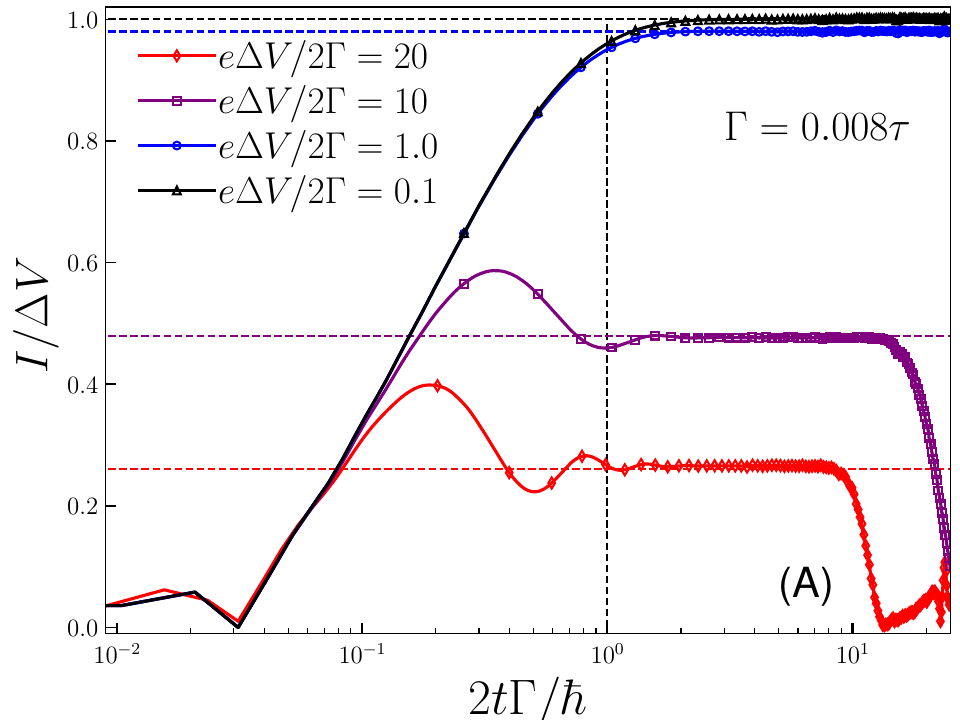}
        &
        \includegraphics[scale=0.418]{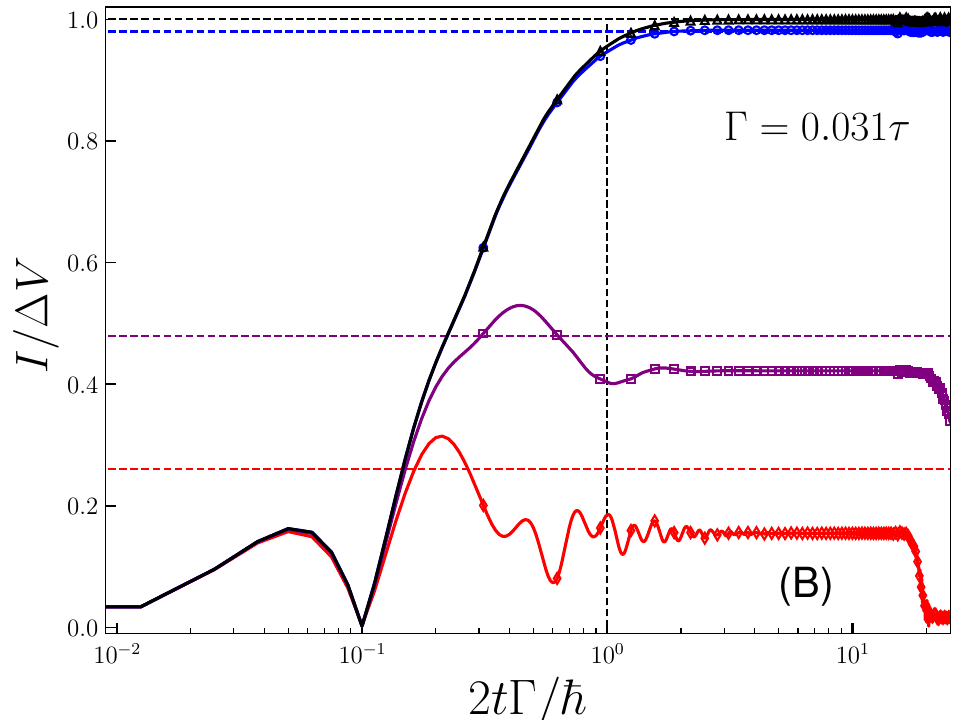}    
        \\
        \includegraphics[scale=0.42]{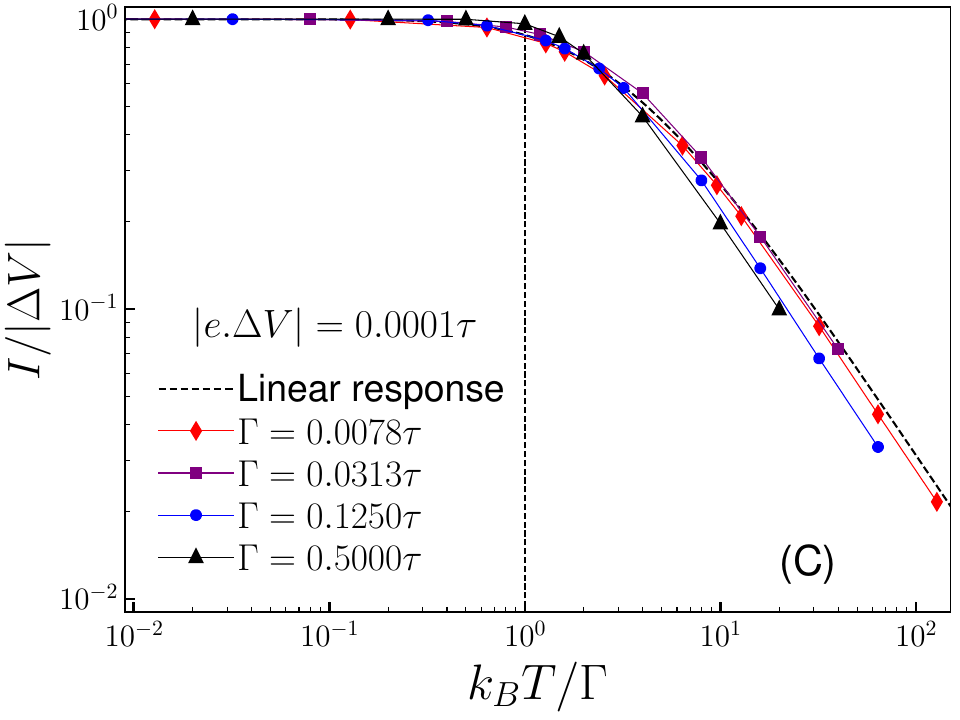} 
        &     
        \includegraphics[scale=0.42]{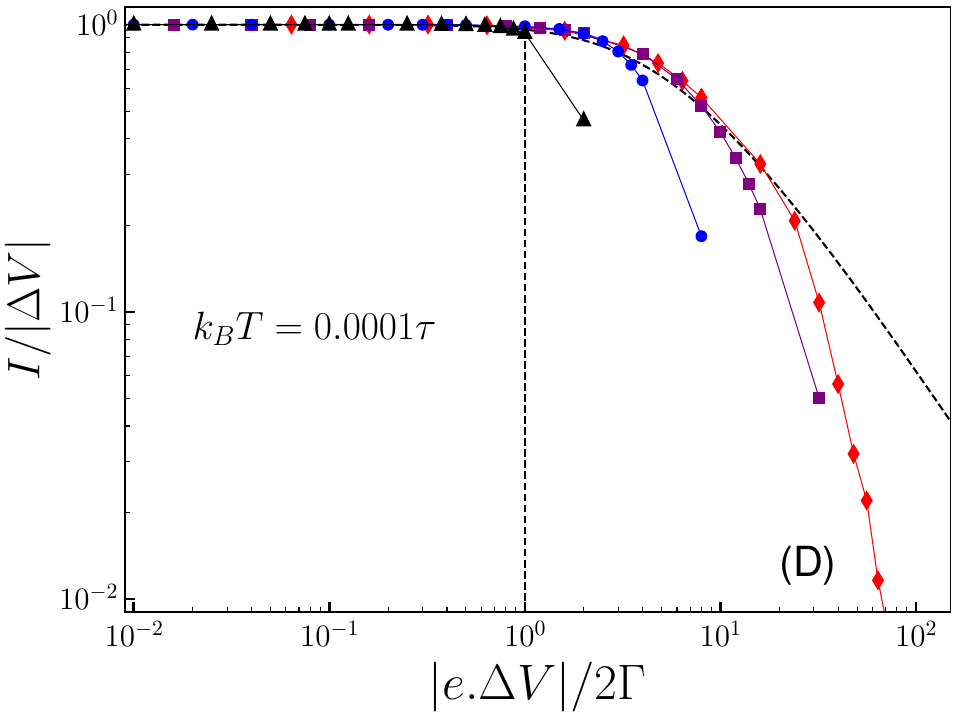}       
        \end{tabular}
    \end{center}
         \caption{Nonequilibrium SET current $I$, divided by $\Delta V$ to observe the nonlinear conductance behavior, in unities of $G_0$. Here, it is fixed $\varepsilon_d = -U/2 = 0$ and $N= 5000$ sites. Time and energies are scaled by $\Gamma$. Top panels (A and B) show the time-dependent current for different values of $e |\Delta V| / 2\Gamma$ (values indicated in the legend of panel A), computed at a fixed $k_B T = 10^{-4} \tau$, for $\Gamma = 0.008 \tau$ (panel A, left) and $\Gamma = 0.031 \tau$ (panel B, right). The dashed lines in the top panels represent the steady-state currents for $\Gamma = 0.008 \tau$, with the corresponding colors, to allow comparison between panels A and B. Bottom panels (C and D) show the steady-state current as a function of $k_B T$ (panel C, left) at fixed $e \Delta V = 10^{-4} \tau$, and as a function of $e \Delta V$ (panel D, right) at fixed $k_B T = 10^{-4} \tau$, for different values of $\Gamma$ (values indicated in the legend of panel C).}
		\label{Current}
\end{figure*}

For example, if the system begins in its ground state and the voltage is increased adiabatically to a non-zero value before being switched to the final bias, it remains in equilibrium up to that point. In such cases, the conductance can be computed using linear response theory (LRT), revealing features such as the Kondo peak splitting shown in Fig. \ref{Kondo_3}. In practice, however, the state of the system during QD manipulation is typically unknown and cannot be assumed to be in equilibrium. Consequently, if the applied voltage is large or introduced non-adiabatically, LRT no longer provides a reliable description.

Given that the ultimate goal is to harness these systems in technological applications --- where fast manipulations and precise control over a wide range of parameters are essential --- the limitations of linear response theory become evident. Meeting these demands requires a deeper, more rigorous investigation of the system, including the explicit computation of transport properties under genuinely nonequilibrium conditions.

\subsection{Nonequilibrium transport}

Computing the nonequilibrium properties of correlated many-body systems is not straightforward, requiring a careful analysis of the relevant energy scales, a task that is far less trivial than in the analogous equilibrium case. However, advances in computational power and time-dependent impurity solver methods have enabled realistic nonequilibrium simulations of these systems \cite{PhysRevB.106.125413}.

For simplicity, let us focus on nonequilibrium cases driven by external perturbations at a fixed temperature. Thus, the nonequilibrium properties of the system can be calculated as
\begin{align}\label{Time_Temp_Obs}
    \langle \hat O \rangle_{\beta,t} \equiv \mathrm{Tr} \left[\hat{U}_t^\dagger \hat O \hat U_t \;\frac{\exp(-\beta H_0)}{\mathcal{Z_\beta}} \right],
\end{align}
where $\hat O$ is the observable operator and $\hat U_t$ is the time-evolution operator.

To verify the limits of accuracy of the LRT, we focus on the non-interacting symmetric case, where $H_0$ is quadratic \cite{diniz2025}.
Figure \ref{Current} shows the nonequilibrium SET current $I$ in response to a sudden introduction of $\Delta V$, normalized by $\Delta V$ to observe the nonlinear behavior. Time and energies are scaled by $\Gamma$\footnote{For the non-interacting symmetric case, this energy scale predominantly determines the properties of the system.}, the numerical calculations are for different values of $e \Delta V/2\Gamma$, for fixed $\Gamma = 0.008 \tau$ (Fig. \ref{Current} A) and $\Gamma = 0.031 \tau$ (Fig. \ref{Current} B). For small values of $e\Delta V/2\Gamma$ (black and blue curves), the scaled conductance is essentially the same for both values of $\Gamma$; however, when $e \Delta V \gg 2 \Gamma$  (red and purple curves), this behavior breaks down, indicating a nonlinear behavior.

Panels C and D show the current\footnote{Specifically, the steady-state current value is chosen, where the current remains constant for extended periods of time.} for several $\Gamma$
as a function of $k_B T$ (Fig. \ref{Current} C) at fixed $e \Delta V = 10^{-4} \tau$, and as a function of $e \Delta V$ (Fig. \ref{Current} D) at fixed $k_B T = 10^{-4} \tau$. Panel C shows good agreement with the LRT prediction, with only a slight deviation appearing for strong hybridization, $\Gamma > 0.125 \tau$ (blue and black curves). However, for any $e \Delta V > 2\Gamma$, Panel D reveals that the system is rapidly driven out of the LRR, described by $I =\Delta V \cdot G(k_BT = e\,|\Delta V|/2)$ from Eq. \eqref{GSET}. Moreover, the larger the value of $\Gamma$, the greater and ealier the deviation occurs.

The non-interacting symmetric case can be used as an analogy to its interacting counterpart (see Fig. \ref{Kondo1}). In the interacting case, the high conductance at low temperatures ($k_B T < T_K$) originates from the effective antiferromagnetic dot-band interaction, which couples the dot to the band low-energy states with an effective hybridization $\Gamma_K \sim k_B T_K$. Similarly, in the non-interacting case, the QD hybridizes with the low-energy states of the band via the hybridization $\Gamma$, allowing high conductivity at low temperatures $k_B T < \Gamma$. Although this analogy is not perfect, one should be aware of potential inaccuracies in predictions made by LRT when $e \Delta V \gg k_B T_K$.

\section{Transport Beyond Linear Regimes}

In the early stages of research on transport through quantum dots (QDs), attention was primarily directed toward the linear regime in relatively simple systems. Today, however, experimental studies frequently probe transport properties beyond this limit. In this section, we highlight both theoretical and experimental developments that, despite the apparent simplicity of the systems involved, continue to reveal rich and unexplored phenomena.

\subsection{Nonlinear conductivity}

Almost twenty years ago, the first signs of nonlinear conductance behavior arising from nonequilibrium effects began to appear. In 2006, Ref. \cite{Paaske2006} reported the observation of a nonequilibrium singlet–triplet Kondo effect in carbon nanotube QDs, where a conductance peak emerges at finite bias when the ground state is a spin singlet and the triplet state becomes accessible through inelastic tunneling. This anomaly, absent at zero bias, was interpreted as a many-body Kondo resonance involving virtual transitions between singlet and triplet states under nonequilibrium conditions, supported by a perturbative renormalization-group analysis. This study demonstrated that Kondo correlations can persist out of equilibrium and contribute to the transport properties of such devices.

Two years later, Ref. \cite{Roch2008} reported a similar phenomenon in a QD based on a C$_{60}$ molecule, further confirming the presence of these nonequilibrium Kondo phenomena in such devices. 

In Ref. \cite{PhysRevB.84.195116}, the author investigated the lineshape of the Kondo resonance in the conductance of spin-$1/2$ magnetic adatoms on decoupling layers at low temperatures, using experimental data obtained from  scanning tunneling microscopy (STM) spectroscopy \cite{Artigo_1011161575444}. As a phenomenological ansatz based on Fano resonances \cite{PhysRevLett.85.2557}, the nonlinear conductance as a function of the applied voltage was expressed as
\begin{eqnarray}\label{ZitkoG}
   G(e.V) &=& c_0 +c_1(1-c_2^2)\mathrm{Im}\mathcal{G}_{\mathrm{K}}\left(\frac{e\,(V-V_0)}{k_B T_K}\right) \nonumber \\ 
        &&+ 2c_2 \mathrm{Re}\mathcal{G}_{\mathrm{K}}\left(\frac{e\,(V-V_0)}{k_B T_K}\right).
\end{eqnarray}
Here, the constants $c_0,c_1$ and $c_2$ can be extracted from the numerical data, as explained in detail in Ref. \cite{PhysRevB.84.195116}; $V_0$ is the bias voltage, $T_K$ is the Kondo temperature, and $\mathcal{G}_{\mathrm{K}}$ is the universal impurity Green’s function for the Kondo model \cite{Zawadzki2018}.

The results obtained from expression \eqref{ZitkoG} show excellent agreement with the experimental data, both at zero bias and under finite bias conditions (in the presence of an external magnetic field), where the Kondo splitting (shown in Fig. \ref{Kondo_3}) is present, confirming the dominance of Kondo physics in this system. Consequently, a Ti adatom on the Cu$_2$N/Cu(100) surface represents a promising platform for investigating nonequilibrium many-body effects through STM tip manipulations.

Subsequently, Ref. \cite{PhysRevB.104.235147} presented a detailed theoretical investigation of the nonlinear transport through a QD, focusing on situations where the couplings to the two leads are asymmetric and the system operates at low temperatures ($T < T_K$), within the Fermi-liquid regime \cite{Nozieres1974_Kondo}. The authors carefully expanded the steady-state current at low energies up to third order in the applied voltage, and obtained the following expression for the conductance:
\begin{eqnarray}\label{G3order}
   \frac{G(e\,V)}{G_0} &=& \sin^2 (\delta) - C_T \left(\frac{\pi T}{T_K}\right)^2 +  C_{V,2} \left(\frac{e\,V}{T_K}\right) \nonumber \\  &&-  C_{V,3} \left(\frac{e\,V}{T_K}\right)^2.
\end{eqnarray}
Here, $\delta$ is the low energy phase shift, and the coefficients can be computed via the corresponding correlation functions \cite{PhysRevB.104.235147}.

This work identified two nonlinear transport coefficients: $C_{V,2}$ and $C_{V,3}$, which encode asymmetry effects, such as finite $\Delta V$ and unequal couplings. $C_{V,2}$ captures the energy shift of the dot’s electronic level induced by asymmetry, while $C_{V,3}$ incorporates contributions from both two-body and three-body Fermi-liquid correlation functions. These coefficients were calculated using NRG calculations, and their behaviors were systematically analyzed. As a consequence, transport measurements in QDs can potentially be used as sensitive probes of many-body Fermi-liquid effects beyond LRR, providing valuable information about the contributions of two- and three-body electronic correlation functions.

In a similar study, Ref. \cite{Hata2021} investigated the effects of three-body electronic correlations on the nonlinear transport properties of low-temperature QDs by combining an extended Fermi-liquid theoretical framework with high-precision measurements on a carbon nanotube QD. Theoretical predictions for the dependence of the three-body contributions, acquired using NRG calculations, are quantitatively validated against the experimental data, providing the first direct experimental evidence of the three-body Fermi-liquid correlations predicted in Ref. \cite{PhysRevB.104.235147}.

Despite the good agreement between theory and experiment reported in Refs. \cite{PhysRevB.84.195116,Hata2021}, where the expressions \eqref{ZitkoG} and \eqref{G3order}, respectively, successfully reproduced the corresponding experimental results, these expressions are not applicable to general cases. In particular, their validity is restricted to $T<T_K$.
While Ref. \cite{PhysRevB.84.195116} does not derive the expression \eqref{ZitkoG} from first principles, Refs. \cite{Hata2021,PhysRevB.104.235147} derive Eq. \eqref{G3order} within a Fermi-liquid description, although it does not fully capture the splitting of the Kondo resonance peak. 

Future works could focus on deriving, from first principles, a more complete expression that generalizes Eq. \eqref{GSET}, incorporating two- and three-body correlation functions, inspired by Refs. \cite{Hata2021,PhysRevB.104.235147}. Such an approach could also be extended to thermopower and spin currents, which are equally important for emerging technologies.  On the experimental side, future studies could use setups similar to those in Refs. \cite{Paaske2006,PhysRevB.84.195116,Roch2008} to investigate nonlinear transport in greater detail, providing valuable data to test and validate theoretical predictions. Additionally, advanced time-dependent simulations of these systems could offer deeper insights into the nature of the excitations responsible for nonlinear effects in such devices \cite{PhysRevB.106.125413}.

\subsection{Violation of the Wiedemann–Franz law}

The Wiedemann–Franz law states that the ratio of a material’s thermal conductivity to its electrical conductivity is proportional to the temperature. This relation can be derived within the LRT formalism \cite{Craven2020}.

Ref. \cite{Majidi2022} reported a clear experimental violation of the WFL in a QD formed within a semiconducting InAs nanowire transistor. Using a precisely tunable SET geometry, the authors measured both the electrical and thermal conductance of the QD under strong confinement at low temperatures. They observed a marked suppression of thermal conductance relative to the value predicted by the law. This suppression was attributed to energy-selective transport: while charge transport remains efficient through discrete energy levels, only a narrow range of electron energies contributes to heat flow, leading to a substantial reduction in thermal conductance. The results are supported by scattering theory and highlight the potential of QDs as thermal conductance filters.

This work indicates the need to extend investigations of thermal transport properties in QDs beyond the LRT as well. Such an extension could be achieved by employing an approach similar to that used for $G(eV)$ in Refs. \cite{Hata2021,PhysRevB.104.235147}, deriving a novel expression for thermal conductivity that incorporates higher-order correlation functions.

\subsection{Direct detection of the Kondo Cloud}

A new approach to directly measure the Kondo cloud length in QDs was proposed in Ref. \cite{PhysRevLett.110.246603}, using an SET-like device with confinement. This method relies on oscillations in conductivity, which appear only when the electrons in the conductor leads are correlated \cite{PhysRevLett.110.246603,PhysRevLett.125.187701}. Using this concept, an experimental setup was built, and oscillations in conductivity as a function of the confinement potential were observed at distances of a few micrometers \cite{Borzenets2020}, which the authors interpreted as direct evidence of the Kondo cloud.

However, despite this promising experimental evidence, there are important considerations regarding the theoretical framework used to analyze these results. In Ref. \cite{Borzenets2020}, the authors applied an alternating small voltage to induce current in the experimental setup, while the theoretical analysis relied on LRT\footnote{Which assumes that the system starts in thermodynamic equilibrium and is subjected to a small voltage. However, as we discussed, a time-dependent voltage can introduce non-zero bias effects and other nonequilibrium behaviors.}. Despite the very interesting observation, the oscillations in conductivity could potentially arise from nonequilibrium effects \cite{PhysRevB.81.100412} or from finite-size effects due to the confinement \cite{PhysRevLett.86.2854}. Therefore, a theoretical study focused on this setup, analyzing the device’s response to an alternating voltage, would greatly help to determine whether these oscillations are indeed a direct consequence of the Kondo cloud.

\subsection{Theoretical methodology}
 
One active line of research focuses on improving the description of QD devices, making them more realistic and directly comparable to experiments by refining their Hamiltonians to incorporate additional effects, such as spin-orbit coupling \cite{PhysRevB.80.041302}, multiple QDs \cite{PRB.73.235337_Karrasch2006,PhysRevB.91.115435,PhysRevB.82.165304,PhysRevLett.121.257701,Wang2022}, and non-metallic leads like superconductors and topological materials \cite{PhysRevB.106.125413,PhysRevLett.121.257701,Gazza2006,Trocha2024}, and other relevant features.

In addition, Ref. \cite{PhysRevLett.119.116801} introduced a generalized conductance formula for nanoscale devices with geometries that violate the Meir–Wingreen proportional-coupling condition. While traditional QD setups assume tunneling amplitudes can be factorized into independent lead and dot contributions, this breaks down in more complex architectures. Using a QD-SET-like coupled to a tunable electronic cavity as a case study, the work investigated how multiple tunneling paths lead to quantum interference, generating asymmetric conductance line shapes and modulations associated with the cavity’s discrete modes. The proposed framework successfully explained experimental observations in QD–cavity systems \cite{PhysRevLett.115.166603}.

However, these works still rely on LRT and are therefore restricted to the linear response regime. To overcome this limitation, two possible strategies have emerged: one is to derive new expressions for the transport properties that include higher-order correlation functions as Refs. \cite{Hata2021,PhysRevB.104.235147}; the other is to adopt time-dependent impurity solvers. Early efforts to generalize impurity solvers to nonequilibrium conditions led to the development of time-dependent DMRG (tDMRG) \cite{PhysRevLett.93.076401} and time-dependent NRG (TDNRG) \cite{PhysRevLett.95.196801}.

The TDNRG \cite{PhysRevLett.95.196801,PhysRevB.98.155107,PhysRevB.89.075118,PhysRevB.74.245113} offers a computationally inexpensive platform to compute nonequilibrium properties of QD-based devices and has proven to be a good alternative for observing the short-time behavior. This method has been improved over the years, but it is still not sufficiently accurate, especially for long-time computations necessary to access steady-state currents \cite{Rosch2012,PhysRevB.93.235159,PhysRevB.98.195138,Picoli}. On the other hand, the tDMRG has demonstrated higher accuracy \cite{Gazza2006,PhysRevLett.93.076401,PhysRevB.79.235336,PhysRevB.111.035445}, though it is computationally demanding and still faces challenges at very long times. Advances in the accuracy of these time-dependent impurity solvers, particularly for long-time behavior, would be extremely valuable for improving our understanding of nonequilibrium transport in these systems.

As an example, the authors of Ref. \cite{PhysRevB.106.125413} investigated nonequilibrium spin transport through a QD coupled to ferromagnetic leads, targeting spintronic applications. Employing a hybrid NRG–tDMRG approach, they analyzed the spin current in response to a suddenly introduced finite bias, for both particle-hole symmetric and asymmetric cases. The study demonstrated that spin polarization suppresses the Kondo conductance peak via an effective exchange field, while applying an external magnetic field can restore the Kondo resonance. Furthermore, the authors systematically mapped how temperature, spin polarization, and external fields jointly govern the nonequilibrium Kondo effect, offering valuable insights for the design of spintronic devices.

Another interesting method for studying the transport in these systems is the use of nonequilibrium Green's functions (NEGF) \cite{Do_2014,PhysRevB.74.085324,Trocha2024,Zhuravel2020,Verma2022}. To demonstrate its usefulness, Ref. \cite{Trocha2024} recently investigated nonequilibrium charge and heat transport in a QD side-coupled to a topological superconductor hosting Majorana zero modes. Using NRGF, the authors derived steady-state expressions for the current and thermocurrent under applied voltage and thermal biases. The coupling to Majorana modes induces zero-bias conductance anomalies and nonmonotonic thermoelectric behavior, including thermocurrent sign reversals. This work proposes transport-based diagnostics for identifying Majorana modes in QD–superconductor hybrid systems.

NEGF methods provide good approximations for small values of $U$, but its perturbative nature makes it unreliable when $U$ becomes large. The TDNRG and tDMRG methods, on the other hand, are non-perturbative and offer accurate results for short-time dynamics. However, these techniques rely on reduced basis approximations and/or logarithmic discretization, which introduce errors in the long-time behavior, particularly in the steady-state regime. Therefore, improving these methods for long-time accuracy is essential for reliably computing steady-state transport properties. Additionally, fast and ultrafast experimental measurements of QD transport would be an excellent way to test and explore the nonequilibrium predictions provided by these theoretical approaches.

\subsection{Fast and ultrafast measurements}

When studying the LRR, the time scale of transport property measurements is not a major issue, since both theoretical and experimental methods can focus on long-time scales. In this regime, typical current measurement times on the order of milliseconds are sufficient. However, nonequilibrium behaviors are much more dependent on time scales. Although their effects still influence long-time transport properties, time-resolved measurements can reveal nonequilibrium phenomena that may remain inaccessible in long-time averaged measurements \cite{Petta2005,Fujisawa2002,Noiri2020,CirianoTejel2021,Stegmann2021,PhysRevResearch.5.013042,Ohkatsu2024}.

In Ref. \cite{Petta2005}, the coherent control of two-electron spin states in a double QD was achieved through rapid electrical tuning of the exchange interaction. To observe this phenomenon, the authors employed radio-frequency measurements techniques, achieving measurement times on the order of nanoseconds. Ref. \cite{Fujisawa2002} studies the allowed and nominally forbidden energy transitions under phonon emission in QDs, based on time-resolved measurements in the nanosecond range. This study shows that forbidden transitions, enabled by spin-flip processes, exhibit relaxation times around 200 $\mu$s, supporting the potential of QDs as candidates for spin-based quantum information storage.

More recently, Refs. \cite{Noiri2020,CirianoTejel2021,Stegmann2021,PhysRevResearch.5.013042,Kerski2023,Ohkatsu2024} have employed similar radio-frequency techniques to achieve fast transport measurements. Ultrafast measurements, capable of capturing phenomena on the picosecond time scale, have also been proposed and discussed for these devices \cite{Petta2005,Shulman2014}. Experimental data with such detailed time dependence would be invaluable for directly comparing observed features with nonequilibrium theoretical predictions. Specially in the transient regime, about which little is currently known.

\subsection{Qubits for quantum computing}

QDs have been studied as potential candidates for qubits in quantum computing \cite{PhysRevA.57.120,PhysRevLett.101.126803,Kandel2021,Fujisawa2002,Ban2018,PhysRevB.70.235317,PhysRevB.97.041416,PhysRevB.99.115424}, as they can confine individual electrons and be easily manipulated, allowing for precise control over their quantum states \cite{RevModPhys.79.1217,Borzenets2020}. For their use as qubits, two key properties of QDs must be ensured: the ability to implement adiabatic protocols to manipulate the devices without losing quantum information \cite{RevModPhys.91.045001,PhysRevLett.101.126803,Kandel2021,PhysRevB.99.115424}, and long coherence times for reliable quantum information storage \cite{PhysRevB.70.235317,Fujisawa2002,Ban2018}.

Adiabatic quantum-state transfer was demonstrated in Ref. \cite{Kandel2021} in chains of semiconductor QD spin qubits via controlled exchange couplings. Single- and two-spin states were transferred over nanoseconds with fidelities exceeding $95\%$\footnote{The fidelity measures how close the quantum state of the system is to the desired state. It is often used to quantify how good the system kept the desired quantum information.}, showing very small loss of information, supporting scalable spin-qubits arrays and future adiabatic quantum computing architectures.

In Ref. \cite{Zajac2018}, a high-fidelity ($98\%$) two-qubit CNOT gate was demonstrated in a silicon QD using electrically tuned exchange interactions and resonant microwave control. Spin-selective manipulation was effectively achieved via electric dipole spin resonance. These results are very promising for the development of electrically controlled, QD-based quantum computing in silicon.

Despite these impressive results, QD-based qubits still lag behind the superconducting qubits \cite{Huang2020} and trapped ions \cite{Monroe2013}, which exhibit longer coherence times, higher fidelities, and simpler control schemes. The challenges in advancing QD-based qubits lie in nanofabrication constraints, charge noise suppression, the implementation of fast and high-fidelity entangling gates, and the development of reliable multi-qubits architectures. As material science and fabrication techniques (particularly in Si/SiGe platforms) continue to improve, QD qubits may yet emerge as a competitive and scalable option for future quantum computing architectures.

\section{Conclusions}

The study of quantum dots beyond the linear regime has uncovered a rich playground for observing new physical phenomena driven by strong correlations and nonequilibrium effects. Notable advances include both the experimental observation and theoretical analysis of nonlinear Kondo resonances in electrical conductivity, where many-body correlations persist even when a finite bias is applied. Furthermore, the experimental violation of the Wiedemann–Franz law in QDs emphasizes the necessity of exploring transport beyond the LRR. The reported direct detection of the Kondo cloud also underscores the potential of QDs as sensitive probes of fundamental phenomena, although a deeper understanding of the interplay between finite-size effects, nonequilibrium dynamics, and quantum confinement remains a significant challenge.

On the theoretical side, recent efforts have focused on refining QD models to incorporate more realistic features. While many of these studies still rely on LRT, advances in non-perturbative, time-dependent methods, such as TDNRG and tDMRG, have offered new insights under nonequilibrium conditions. Although these techniques currently face limitations in capturing long-time dynamics, they remain crucial tools for exploring the behavior of strongly correlated systems and are continuously undergoing improvements.

Complementarily, fast and ultrafast experimental techniques --- capable of probing dynamics on nanosecond and even picosecond timescales --- have opened exciting new avenues for investigating real-time transport phenomena, with promising applications in spintronics and quantum information. As QD-based qubits continue to advance, overcoming key challenges related to coherence, precise control, and scalability will be essential for establishing QDs as viable building blocks in future quantum technologies.

\textbf{Acknowledgments} - The authors thank E. Vernek for fruitful discussions. We gratefully acknowledge the support and computational resources provided by the Centre for Mathematical Sciences Applied to Industry (CeMEAI), funded by São Paulo Research Foundation (FAPESP) under the grant No. 2013/07375-0. VVF is supported by FAPESP (Grants
No. 2021/06744-8, 2025/02935-4, and 2024/00998-6) and
by the National Council for Scientific and Technological
Development (CNPq), Grants No. 403890/2021-7 and
306301/2022-9.
~
\renewcommand*{\bibfont}{\footnotesize}
\bibliography{bibliography}

\end{document}